\def\vs{\vskip.3cm}

\def\+{{(+)}}  \def\-{ {(-)} }   \def\0{ {(0)} }
\def\1{ {(1)} }  \def\2{ {(2)} }

                      \def\d{\delta}

\def\d{\delta}
\def\sq{Q\kern-6pt/}
\def\sQ{Q\kern-12pt\nearrow}
\documentclass[11pt,eqs]{article}

\usepackage{latexsym}
\usepackage{amssymb}

\def\be{\begin{equation}}             \def\ee{\end{equation}}
\def\ba{\begin{array}{rcl}}           \def\ea{\end{array}}
\def\beqa{\begin{eqnarray} }          \def\eeqa{\end{eqnarray} }
\def\beqalign{\begin{eqalign}}        \def\eeqalign{\end{eqalign}}
             
\def\bsubeq{\begin{subequations}}     \def\esubeq{\end{subequations}}
\def\bitem{\begin{itemize}}           \def\eitem{\end{itemize}}

\def\DJ{\leavevmode\setbox0=\hbox{D}\kern0pt
 \rlap{\kern.04em\raise.188\ht0\hbox{-}}D}
\def\dj{\leavevmode\setbox0=\hbox{d}\kern0pt
 \rlap{\kern.215em\raise.46\ht0\hbox{-}}d}

\newcommand{\bd}{\begin{displaymath}}
\newcommand{\ed}{\end{displaymath}}

\begin{document}

\title{ Fermionic T-duality in fermionic double space
\thanks{Work supported in part by the Serbian Ministry of Education, Science and Technological Development, under contract No. 171031.}}
\author{B. Nikoli\'c \thanks{e-mail address: bnikolic@ipb.ac.rs} and B. Sazdovi\'c
\thanks{e-mail address: sazdovic@ipb.ac.rs}\\
       {\it Institute of Physics}\\{\it University of Belgrade}\\{\it P.O.Box 57, 11001 Belgrade, Serbia}}
\maketitle

\begin{abstract}

In this article we offer the interpretation of the fermionic T-duality of the type II superstring theory in double space.
We generalize the idea of double space doubling the fermionic sector of the superspace. In such doubled space
fermionic T-duality is represented as permutation of the fermionic coordinates $\theta^\alpha$ and $\bar\theta^\alpha$ with
the corresponding fermionic T-dual ones, $\vartheta_\alpha$ and $\bar\vartheta_\alpha$, respectively. Demanding that T-dual transformation law
has the same form as initial one, we obtain the known form of the fermionic T-dual NS-R i R-R background fields. Fermionic T-dual NS-NS background fields are obtained under some assumptions.
We conclude that only symmetric part of R-R field strength and symmetric part of its fermionic T-dual contribute to the fermionic T-duality transformation of dilaton field and analyze the dilaton field in fermionic double space.
As a model we use the ghost free action of type II superstring in pure spinor formulation in approximation
of constant background fields up to the quadratic terms.

\end{abstract}
\vs

\section{Introduction}
\setcounter{equation}{0}

Two theories
T-dual to one another can be viewed as being physically identical  \cite{jopol,rabin}. T-duality presents an important tool which shows the equivalence of different geometries and topologies.
The useful T-duality procedure was first introduced by
Buscher \cite{buscher}.

Mathematical realization of T-duality is given by Buscher T-dualization procedure \cite{buscher}, which is considered as standard one. There are also other frameworks
in which we can represent T-dualization which should agree with the Buscher procedure.
It is double space formalism which was the subject of the articles about twenty years
ago \cite{Duff,AAT1,AAT2,WS1,WS2}. Double space is spanned by coordinates $Z^M=(x^\mu\;\;\;y_\mu)^T$ $(\mu=0,1,2,\dots,D-1)$,
where $x^\mu$ and $y_\mu$ are the coordinates of the $D$-dimensional initial and T-dual space-time, respectively.
Interest for this subject emerged recently with papers \cite{Hull,Hull2,berman,negeom,hohmz},
where T-duality along some subset of $d$ coordinates is considered as $O(d,d)$ symmetry transformation and \cite{sazdam,sazda},
where it is considered as permutation of $d$ initial with corresponding $d$ T-dual coordinates.

Until recently only T-duality along bosonic coordinates has been considered. Analyzing the gluon scattering amplitudes in $N=4$ super Yang-Mills theory, a new kind of T-dual symmetry, fermionic T-duality, was discovered \cite{ferdual,MW1}. It is a part of
the dual superconformal symmetry which should be connected to integrability and it is valid just at string tree level.
Mathematically, fermionic T-duality is realized within the same procedure as bosonic one, except that dualization is performed along
fermionic variables. So, it can be considered as a generalization of Buscher T-duality.
Fermionic T-duality consists in certain non-local redefinitions of the fermionic variables of the superstring mapping a supersymmetric background to another supersymmetric background. In Refs.\cite{ferdual,MW1} it was shown
that fermionic T-duality maps gluon scattering amplitudes in the original theory to an object very close to Wilson loops in the dual one. Calculation of gluon scattering amplitudes in the initial theory is equivalent to the calculation of Wilson loops in fermionic T-dual
theory. Generalizing the idea of double space to the fermionic case we would get fermionic double space in which fermionic T-duality is a symmetry \cite{Siegel1} which exchanges scattering amplitudes and Wilson loops. Fermionic double space can be also successfully
applied in random lattice \cite{Siegel2}, where doubling of the supercoordinate was done. Relation between fermionic T-duality and open string noncommutativity was considered in Ref.\cite{nashi}.

Let us explain our motivation for fermionic T-duality.
It is well known that T-duality is important feature  in understanding the M-theory. In fact, five consistent  superstring theories are connected by web of T and S dualities. We are going to pay attention to the T-duality, hoping that
S-duality (which can be understood as transformation of dilaton background field also) can be later  successfully  incorporated  into our procedure.
If we start with arbitrary (of five consistent  superstring) theory and find all corresponding T-dual theories we can achieve any of other four consistent  superstring theories.
But to  obtain formulation of M-theory it is not enough. We must construct one theory which contains the initial theory and all corresponding T-dual ones.

In the bosonic case (which is substantially simpler that supersymmetric  one)   we have succeeded to realize such program. In Refs.\cite{sazdam,sazda}  we doubled all bosonic coordinates and showed that such theory contained the initial and all corresponding T-dual theories. We can connect arbitrary two of these theories just replacing some initial coordinates $x^a$ with corresponding   T-dual ones $y_a$. This is equivalent with T-dualization along coordinates $x^a$.  So, introducing double space T-duality ceases to be  transformation which connects two physically equivalent theories but it becomes  symmetry transformation in extended space
with respect to  permutation group.
We proved this in the bosonic string case both for constant and for weakly curved background with  linear  dependence   on coordinates.

Unfortunately, this is not enough for construction of M-theory, because the T-duality for superstrings is much more complicated then in the bosonic case \cite{rbgpjt}.
In Ref.\cite{bosdouble} we have tried to extend such approach to the type II theories. In fact, doubling all bosonic coordinates we have unified types IIA, IIB as well as type $II^\star$ \cite{timelike}
(obtained by T-dualization along time-like direction)  theories. There is an incompleteness in such approach. Doubling all bosonic coordinates, by simple permutations of initial with corresponding T-dual coordinates, we obtained all T-dual background fields except T-dual R-R field strength $F^{\alpha \beta}$. To obtain
${}_a F^{\alpha \beta}$ (the field strength after T-dualization along coordinates $x^a$) we  need to introduce some additional assumptions. The explanation is that  R-R field strength $F^{\alpha\beta}$ appears  coupled
with fermionic momenta $\pi_\alpha$ and $\bar\pi_\alpha$ along which we did not performed T-dualization and consequently we did not double these variables. It is an analogue of $ij$-term in approach of Refs.\cite{Hull,Hull2} where $x^i$ coordinates are not doubled.

Therefore, in the first step of our approach to the formulation of M-theory (unification of types II theories)  we must include  T-dualization along fermionic variables ($\pi_\alpha$ and $\bar\pi_\alpha$ in particular case).
It means that we should doubled these fermionic variables, also. The present article  represents a necessary step for understanding T-dualization along all fermionic coordinates in fermionic double space. We expect that final
step in construction of M-theory will be unification of all theories obtained after T-dualization along all bosonic and all fermionic variables \cite{Siegel1,Siegel2}. In that case we should  double all coordinates in superspace, anticipating that  some
super permutation will connect arbitrary two of our five consistent super symmetric string theories.

In this article we are going to double
fermionic sector of type II theories adding to the coordinates $\theta^\alpha$ and $\bar\theta^\alpha$ their fermionic T-duals, $\vartheta_\alpha$ and $\bar\vartheta_\alpha$, where index $\alpha$ counts independent real components of
the spinors, $\alpha=1,2,\dots,16$. Rewriting T-dual transformation
laws in terms of the double coordinates, $\Theta^A=(\theta^\alpha,\vartheta_\alpha)$ and $\bar\Theta^A=(\bar\theta^\alpha,\bar\vartheta_\alpha)$, we define the "fermionic generalized metric" ${\cal F}_{AB}$ and the generalized currents $\bar{\cal J}_{+ A}$ and ${\cal J}_{-A}$.
The permutation matrix ${\cal T}^A{}_B$ exchanges $\bar\theta^\alpha$ and $\theta^\alpha$ with their T-dual partners, $\bar\vartheta_\alpha$ and $\vartheta_\alpha$, respectively. From the requirement that fermionic T-dual coordinates, ${}^\star \Theta^A={\cal T}^A{}_B \Theta^B$ and ${}^\star\bar\Theta^A={\cal T}^A{}_B \bar\Theta^B$, have the same transformation law as initial ones, $\Theta^A$ and $\bar\Theta^A$, we obtain the expressions for
fermionic T-dual
generalized metric, ${}^\star {\cal F}_{AB}=({\cal T}{\cal F}{\cal T})_{AB}$, and T-dual currents, ${}^\star \bar{\cal J}_{+ A}={\cal T}_A{}^B \bar{\cal J}_{+ B}$ and ${}^\star {\cal J}_{- A}={\cal T}_A{}^B {\cal J}_{- B}$, in terms of the initial ones. These expressions
produce the expression for fermionic T-dual NS-R fields and R-R field strength. Expressions for fermionic T-dual metric and Kalb-Ramond field are obtained separately under some assumptions.
We conclude that only symmetric part of R-R field strength, $F_s^{\alpha\beta}=\frac{1}{2}(F^{\alpha\beta}+F^{\beta\alpha})$, and symmetric part of its fermionic
T-dual, ${}^\star F_{\alpha\beta}^s=\frac{1}{2}({}^\star F_{\alpha\beta}+{}^\star F_{\beta\alpha})$, give contribution to the dilaton field transformation under fermionic T-duality. We also investigate the dilaton field in double space.

\section{Type II superstring and fermionic T-duality}
\setcounter{equation}{0}

In this section we will introduce the action of type II
superstring theory in pure spinor formulation and perform fermionic T-duality \cite{ferdual,MW1,nashi} using fermionic analogue of Buscher rules \cite{buscher}.

\subsection{Action and supergravity constraints}

In this manuscript we use the action of type II superstring theory in pure spinor formulation \cite{berko} up to the quadratic terms with constant background fields. Here we will derive the final form of the action which will be exploited in the further analysis.
It corresponds to the actions used in Refs.\cite{susyNC,BNBSPLB, bnbsjhep, bnbsnpb}.

The sigma model action for type II superstring of
Ref.\cite{verteks} is of the form
\begin{equation}\label{eq:dejstvo1}
S=S_0+V_{SG}\, ,
\end{equation}
where $S_0$ is the action in the flat background
\begin{equation}
S_0=\int_\Sigma d^2\xi \left( \frac{\kappa}{2}\eta^{mn}\eta_{\mu\nu}\partial_m x^\mu \partial_n x^\nu-\pi_\alpha \partial_{-} \theta^\alpha+\partial_+ \bar\theta^\alpha \bar\pi_\alpha\right)+S_\lambda+S_{\bar\lambda}\, ,
\end{equation}
and it is deformed by integrated form of the massless type II supergravity
vertex operator
\begin{equation}\label{eq:vsg}
V_{SG}=\int_\Sigma d^2 \xi (X^T)^M A_{MN}\bar X^N\, .
\end{equation}
The vectors $X^M$ and $X^N$ are defined as
\begin{equation}
X^M=\left(
\begin{array}{c}
\partial_+ \theta^\alpha\\\Pi_+^\mu\\d_\alpha\\\frac{1}{2}N_+^{\mu\nu}
\end{array}\right)\, ,\quad \bar X^M=\left(
\begin{array}{c}
\partial_-\bar\theta^\alpha\\\Pi_-^\mu\\\bar d_\alpha\\\frac{1}{2}\bar N_-^{\mu\nu}
\end{array}\right),
\end{equation}
and supermatrix $A_{MN}$ is of the form
\begin{equation}\label{eq:Amn}
A_{MN}=\left(\begin{array}{cccc}
A_{\alpha\beta} & A_{\alpha\nu} & E_\alpha{}^\beta & \Omega_{\alpha,\mu\nu}\\
A_{\mu\beta} & A_{\mu\nu} & \bar E_\mu^\beta & \Omega_{\mu,\nu\rho}\\
E^\alpha{}_\beta & E^\alpha_\nu & {\rm P}^{\alpha\beta} & C^\alpha{}_{\mu\nu}\\
\Omega_{\mu\nu,\beta} & \Omega_{\mu\nu,\rho} & \bar C_{\mu\nu}{}^\beta & S_{\mu\nu,\rho\sigma}
\end{array}\right)\, ,
\end{equation}
where notation and definitions are taken from Ref.\cite{verteks}. The actions for pure spinors, $S_\lambda$ and $S_{\bar\lambda}$, are free field actions and fully decoupled from the rest of action $S_0$.
The world sheet $\Sigma$ is parameterized by
$\xi^m=(\xi^0=\tau\, ,\xi^1=\sigma)$ and
$\partial_\pm=\partial_\tau\pm\partial_\sigma$. Bosonic part of superspace is
spanned by coordinates $x^\mu$ ($\mu=0,1,2,\dots,9$), while
the fermionic one is spanned by $\theta^\alpha$ and $\bar\theta^{\alpha}$
$(\alpha=1,2,\dots,16)$. The variables $\pi_\alpha$ and $\bar
\pi_{\alpha}$ are canonically conjugated momenta to
$\theta^\alpha$ and $\bar\theta^\alpha$, respectively. All spinors
are Majorana-Weyl ones, which means that each of them has 16 independent real components. Matrix with superfields generally depends on $x^\mu$,
$\theta^\alpha$ and $\bar\theta^\alpha$.

The superfields $A_{\mu\nu}$, $\bar E_\mu{}^\alpha$, $E^\alpha{}_\mu$ and ${\rm P}^{\alpha\beta}$ are known as physical superfields, while the fields given in the first column and first row are auxiliary superfields because they can be expressed in
terms of the physical ones \cite{verteks}. The rest ones, $\Omega_{\mu,\nu\rho}(\Omega_{\mu\nu,\rho})$, $C^\alpha{}_{\mu\nu}(\bar C_{\mu\nu}{}^\alpha)$ and $S_{\mu\nu,\rho\sigma}$, are curvatures (field strengths) for physical superfields.

The expanded form of the vertex operator (\ref{eq:vsg}) is \cite{verteks}
\begin{eqnarray}
V_{SG}&=&\int d^2\xi \left[\partial_+ \theta^\alpha A_{\alpha\beta}\partial_-\bar\theta^\beta+\partial_+ \theta^\alpha A_{\alpha\mu}\Pi_-^\mu+\Pi_+^\mu A_{\mu\alpha}\partial_-\bar\theta^\alpha+\Pi_+^\mu A_{\mu\nu}\Pi_-^\nu\right.\nonumber\\
&+& d_\alpha E^\alpha{}_\beta \partial_-\bar\theta^\beta+d_\alpha E^\alpha{}_\mu \Pi_-^\mu+\partial_+ \theta^\alpha E_\alpha{}^\beta \bar d_\beta+\Pi_+^\mu E_\mu{}^\beta \bar d_\beta+d_\alpha {\rm P}^{\alpha\beta}\bar d_\beta\nonumber\\
&+& \frac{1}{2}N_+^{\mu\nu} \Omega_{\mu\nu,\beta} \partial_- \bar\theta^\beta+\frac{1}{2}N_+^{\mu\nu} \Omega_{\mu\nu,\rho} \Pi_-^\rho+\frac{1}{2}\partial_+\theta^\alpha \Omega_{\alpha,\mu\nu}\bar N_-^{\mu\nu}+\frac{1}{2}\Pi_+^\mu \Omega_{\mu,\nu\rho}\bar N_-^{\nu\rho}\nonumber\\
&+&\left. \frac{1}{2}N_+^{\mu\nu}\bar C_{\mu\nu}{}^\beta \bar d_\beta+\frac{1}{2}d_\alpha C^\alpha{}_{\mu\nu} \bar N_-^{\mu\nu}+\frac{1}{4}N_+^{\mu\nu} S_{\mu\nu,\rho\sigma}\bar N_-^{\rho\sigma}\right]\, .
\end{eqnarray}
The supergravity constraints
are the conditions obtained as a consequence of nilpotency and (anti)holomorphicity of BRST operators $Q=\int \lambda^\alpha d_\alpha$ and $\bar Q=\int \bar\lambda^\alpha \bar d_\alpha$, where
$\lambda^\alpha$ and $\bar\lambda^\alpha$ are pure spinors and $d_\alpha$ and $\bar d_\alpha$ are independent variables.
Let us discuss the choice of background fields satisfying superspace equations of motion in the context of supergravity constraints which are explained in details for pure spinor formalism in Refs.\cite{nbph,verteks}.

In order to implement T-duality many restrictions should  be imposed. For example, in bosonic case one should assume the existence of Killing vectors,
which in fact means background fields independence on corresponding suitably selected coordinates. The idea is to avoid dependence on the coordinate $x^\mu$ and allow only
dependence on the $\sigma$ and $\tau$ derivatives of the coordinates, ${\dot x}^\mu$ and $x^{\prime \mu}$.
The case with explicit dependence on the coordinate requires particular attention and has been considered in Ref.\cite{DS}.
Similar simplifications must be imposed in consideration of the non-commutativity of the coordinates \cite{NS,DS}.

A similar situation occurs in the supersymmetric case. In order to perform fermionic T-duality we must
avoid explicit dependence of background fields on the fermionic coordinates $\theta^\alpha$ and $\bar\theta^\alpha$ (fermionic coordinates are Killing spinors) and
allow only dependence on the $\sigma$ and $\tau$ derivatives of these  coordinates.
Assumption of existence of Killing spinors produces that the auxiliary  superfields should be taken to be zero what can be seen from Eq.(5.5) of Ref.\cite{verteks}.

The right-hand side of the equations of motion for background fields (see for example \cite{dufftasi}) is energy-momentum tensor which is generally square of field strengths.
In our case physical superfields $G_{\mu\nu}$, $B_{\mu\nu}$, $\Phi$, $\Psi_\mu^\alpha$ and $\bar\Psi^\alpha_\mu$ are constant (do not depend on $x^\mu$, $\theta^\alpha$,$\bar\theta^\alpha$)
and corresponding field strengths, $\Omega_{\mu,\nu\rho}(\Omega_{\mu\nu,\rho})$, $C^\alpha{}_{\mu\nu}(\bar C_{\mu\nu}{}^\alpha)$ and $S_{\mu\nu,\rho\sigma}$, are zero. The only nontrivial contribution
of the quadratic terms in equations of motion
comes from constant field strength ${\rm P}^{\alpha\beta}$.
It can induce back-reaction to the background fields. In order to analyze this issue we will use relations from Eq.(3.6) of Ref.\cite{verteks}
labeled by $(\frac{1}{2},\frac{3}{2},\frac{3}{2})$
\begin{equation}\label{eq:red10}
D_\alpha {\rm P}^{\beta\gamma}-\frac{1}{4}(\Gamma^{\mu\nu})_\alpha{}^\beta \bar C_{\mu\nu}{}^\gamma=0\, ,\quad \bar D_\alpha {\rm P}^{\beta\gamma}-\frac{1}{4}(\Gamma^{\mu\nu})_\alpha{}^\gamma C^\beta{}_{\mu\nu}=0\, ,
\end{equation}
in which derivative of ${\rm P}^{\alpha\beta}$ appears. Here
\begin{equation}
D_\alpha=\frac{\partial}{\partial \theta^\alpha}+\frac{1}{2}(\Gamma^\mu \theta)_\alpha \frac{\partial}{\partial x^\mu}\, ,\quad \bar D_\alpha=\frac{\partial}{\partial \bar\theta^\alpha}+\frac{1}{2}(\Gamma^\mu \bar\theta)_\alpha \frac{\partial}{\partial x^\mu}\, ,
\end{equation}
are superspace covariant derivatives
and $C^\alpha{}_{\mu\nu}$ and $\bar C_{\mu\nu}{}^\alpha$ are field strengths for gravitino fields  $\Psi^\alpha_\mu$ and $\bar\Psi_\mu^\alpha$, respectively.
In order to perform fermionic T-dualization along all fermionic directions, $\theta^\alpha$ and $\bar\theta^\alpha$, we assume that they are Killing spinors which means
\begin{equation}
\frac{\partial {\rm P}^{\beta\gamma}}{\partial \theta^\alpha}=\frac{\partial {\rm P}^{\beta\gamma}}{\partial \bar\theta^\alpha}=0\, .
\end{equation}
Taking into account that gravitino fields, $\Psi^\alpha_\mu$ and $\bar\Psi^\alpha_\mu$, are constant (corresponding field strengths are zero), from the equations (\ref{eq:red10}) it follows
\begin{equation}\label{eq:con1}
(\Gamma^\mu)_{\alpha\delta} \partial_\mu {\rm P}^{\beta\gamma}=0\, .
\end{equation}
Note that this is more general case than equation of motion for R-R field strength, $(\Gamma^\mu)_{\alpha\beta}\partial_\mu {\rm P}^{\beta\gamma}=0$, given in Eq.(3.11) of Ref.\cite{verteks} where there is summation over spinor indices.
Our choice of constant ${\rm P}^{\alpha\beta}$ is consistent with this condition. 
It is known fact that even constant R-R field strength produces back-reaction on background fields. 
In order to cancel non-quadratic terms originating from back-reaction, the constant R-R field strength must satisfy additional conditions - $AdS_5 \times S_5$ coset geometry or self-duality condition.

Taking into
account these assumptions there exists solution
\begin{equation}
\Pi_\pm^\mu\to \partial_{\pm} x^\mu\, ,\quad d_\alpha\to \pi_\alpha\, ,\quad \bar d_\alpha\to \bar\pi_\alpha\, ,
\end{equation}
and only nontrivial superfields take the form
\begin{equation}
A_{\mu\nu}=\kappa(\frac{1}{2}g_{\mu\nu}+B_{\mu\nu})\, ,\quad E^\alpha_\nu=-\Psi^\alpha_\nu\, ,\quad \bar E_\mu^\alpha=\bar\Psi_\mu^\alpha\, ,\quad {\rm P}^{\alpha\beta}=\frac{2}{\kappa}P^{\alpha\beta}=\frac{2}{\kappa}e^{\frac{\Phi}{2}}F^{\alpha\beta}\, ,
\end{equation}
where $g_{\mu\nu}$ is symmetric and $B_{\mu\nu}$ is antisymmetric
tensor.

The final form of the vertex operator under these assumptions is
\begin{eqnarray}
V_{SG}=\int_\Sigma d^2\xi \left[
\kappa(\frac{1}{2}g_{\mu\nu}+B_{\mu\nu})\partial_+x^\mu\partial_-x^\nu-\pi_\alpha
\Psi^\alpha_\mu \partial_-x^\mu+\partial_+ x^\mu
\bar\Psi^\alpha_\mu\bar\pi_\alpha+\frac{2}{\kappa}\pi_\alpha
P^{\alpha\beta}\pi_\beta\right]\, .
\end{eqnarray}
Consequently, the action $S$ is of the form
\begin{eqnarray}\label{eq:SB}
&{}&S=\kappa \int_\Sigma d^2\xi \left[\partial_{+}x^\mu
\Pi_{+\mu\nu}\partial_- x^\nu+\frac{1}{4\pi\kappa}\Phi R^{(2)}\right] \\&+&\int_\Sigma d^2 \xi \left[
-\pi_\alpha
\partial_-(\theta^\alpha+\Psi^\alpha_\mu
x^\mu)+\partial_+(\bar\theta^{\alpha}+\bar \Psi^{\alpha}_\mu
x^\mu)\bar\pi_{\alpha}+\frac{2}{\kappa}\pi_\alpha P^{\alpha
\beta}\bar \pi_{\beta}\right ]\, ,\nonumber
\end{eqnarray}
where $G_{\mu\nu}=\eta_{\mu\nu}+g_{\mu\nu}$ and
\begin{equation}\label{eq:pipm}
\Pi_{\pm \mu\nu}=B_{\mu\nu}\pm \frac{1}{2}G_{\mu\nu}\, .
\end{equation}
All terms containing pure spinors vanished because curvatures are zero under our assumption that
physical superfields are constant. Actions $S_\lambda$ and $S_{\bar\lambda}$ are fully decoupled from the rest action and can be neglected in the further analysis. The action, in its final form, is  ghost independent.

Here we work both with type IIA and type IIB superstring theory. The difference is in the chirality of NS-R background fields and content of R-R sector. In NS-R sector there are two gravitino fields
$\Psi^\alpha_\mu$ and $\bar\Psi^\alpha_\mu$ which are Majorana-Weyl spinors of the opposite chirality in type IIA and same chirality in type IIB theory. The same feature
stands for the pairs of spinors $(\theta^\alpha,\bar\theta^\alpha)$ and $(\pi_\alpha,\bar\pi_\alpha)$. The R-R field strength $F^{\alpha\beta}$ is expressed in terms of the antisymmetric tensors $F_{(k)}=F_{\mu_1\mu_2\dots\mu_k}$ \cite{jopol}
\begin{equation}\label{eq:RRpolje}
F^{\alpha\beta}=\sum_{k=0}^D
\frac{1}{k!}F_{(k)}(C\Gamma_{(k)})^{\alpha\beta}\, , \quad \left[
\Gamma_{(k)}^{\alpha\beta}=(\Gamma^{[\mu_1\dots
\mu_k]})^{\alpha\beta}\right]
\end{equation}
where
\begin{equation}
\Gamma^{[\mu_1 \mu_2\dots \mu_k]}\equiv \Gamma^{[\mu_1}\Gamma^{\mu_2}\dots \Gamma^{\mu_k]}\, ,
\end{equation}
is basis of completely antisymmetrized product of gamma matrices and $C$ is charge conjugation operator. For more technical details regarding gamma matrices see the first reference in \cite{jopol}.

R-R field strength satisfies the chirality condition $\Gamma^{11} F=\pm F\Gamma^{11}$, where $\Gamma^{11}$ is a product of gamma matrices in $D=10$ dimensional space-time. The sign $+$ corresponds to
type IIA while sign $-$ corresponds to type IIB superstring theory. Consequently, type IIA theory
contains only even rank tensors $F_{(k)}$, while type IIB contains only odd rank tensors. For type IIA the independent tensors are $F_{(0)}$, $F_{(2)}$ and $F_{(4)}$, while
independent tensors for type IIB are $F_{(1)}$, $F_{(3)}$ and
self-dual part of $F_{(5)}$.

\subsection{Fixing the chiral gauge invariance}

The fermionic part of the action (\ref{eq:SB}) has the form of the first order theory. We want to eliminate the fermionic momenta and work with the action expressed in terms of coordinates and their derivatives. So,
on the equations of motion for fermionic momenta $\pi_\alpha$ and $\bar\pi_\alpha$,
\begin{equation}\label{eq:impulsi}
\pi_\alpha=-\frac{\kappa}{2}\partial_+\left(\bar\theta^\beta+\bar\Psi^\beta_\mu x^\mu\right)(P^{-1})_{\beta\alpha}\, ,\quad \bar\pi_\alpha=\frac{\kappa}{2}(P^{-1})_{\alpha\beta}\partial_-\left(\theta^\beta+\Psi^\beta_\mu x^\mu\right)\, ,
\end{equation}
the action gets the form
\begin{eqnarray}\label{eq:lcdejstvo}
&{}&S(\partial_\pm x, \partial_- \theta, \partial_+
\bar\theta)=\kappa \int_\Sigma d^2\xi \partial_+ x^\mu
\Pi_{+\mu\nu}\partial_-x^\nu\nonumber+\frac{1}{4\pi}\int_\Sigma d^2\xi \Phi R^{(2)}\\&+&\frac{\kappa}{2} \int_\Sigma d^2\xi \partial_+\left(\bar\theta^\alpha+\bar\Psi^\alpha_\mu x^\mu\right)
(P^{-1})_{\alpha\beta}\partial_-\left(\theta^\beta+\Psi^\beta_\nu x^\nu\right)\nonumber\\
&=& \kappa \int_\Sigma d^2 \xi \partial_+ x^\mu \left[\Pi_{+\mu\nu}+\frac{1}{2}\bar\Psi^\alpha_\mu (P^{-1})_{\alpha\beta}\Psi^\beta_\nu\right]\partial_- x^\nu+\frac{1}{4\pi}\int_\Sigma d^2\xi \Phi R^{(2)}\\
&+&\frac{\kappa}{2} \int_\Sigma d^2 \xi \left[\partial_+ \bar\theta^\alpha (P^{-1})_{\alpha\beta}\partial_-\theta^\beta +\partial_+ \bar\theta^\alpha (P^{-1}\Psi)_{\alpha\mu}\partial_-x^\nu+\partial_+x^\mu (\bar\Psi P^{-1})_{\mu\alpha}\partial_- \theta^\alpha\right]\nonumber\, .
\end{eqnarray}

In the above action  $\theta^\alpha$ appears only in the form $\partial_-\theta^\alpha$ and $\bar\theta^\alpha$ in the form $\partial_+ \bar\theta^\alpha$. This means that the theory has a local
symmetry
\begin{equation}
\delta \theta^\alpha=\varepsilon^\alpha(\sigma^+)\, ,\quad \delta \bar\theta^\alpha=\bar\varepsilon^\alpha(\sigma^-)\, ,\quad (\sigma^\pm=\tau\pm\sigma ) \, .
\end{equation}
We will treat this symmetry within BRST formalism. The corresponding BRST transformations are
\begin{equation}
s \theta^\alpha=c^\alpha(\sigma^+)\, ,\quad s \bar\theta^\alpha=\bar c^\alpha(\sigma^-)\, ,
\end{equation}
where for each gauge parameter $\varepsilon^\alpha(\sigma^+)$ and $\bar\varepsilon^\alpha(\sigma^-)$ we introduced the ghost fields $c^\alpha(\sigma^+)$ and $\bar c^\alpha(\sigma^-)$, respectively.
Here $s$ is BRST nilpotent operator.

To fix
gauge freedom we introduce gauge fermion with ghost number $-1$
\begin{equation}
\Psi=\frac{\kappa}{2}\int d^2\xi \left[\bar C_{\alpha}\left(\partial_+\theta^\alpha+\frac{\alpha^{\alpha\beta}}{2}b_{+\beta}\right)+\left(\partial_- \bar\theta^\alpha+\frac{1}{2}\bar b_{-\beta}\alpha^{\beta\alpha}\right)C_{\alpha}\right]\, ,
\end{equation}
where $\alpha^{\alpha\beta}$ is arbitrary non singular matrix,  $\bar C_{\alpha}$ and  $C_{\alpha}$ are antighost fields, while $b_{+\alpha}$ and $\bar b_{-\alpha}$ are Nakanishi-Lautrup auxillary fields which satisfy
\begin{equation}
 s C_\alpha=b_{+\alpha}\, ,\quad s\bar C_\alpha=\bar b_{-\alpha}\, ,\quad s b_{+\alpha}=0\quad s \bar b_{-\alpha}=0\, .
\end{equation}
BRST transformation of gauge fermion $\Psi$ produces the gauge fixed and Fadeev-Popov action
\begin{eqnarray}
&&s \Psi=S_{gf}+S_{FP}\nonumber \, ,\\
&& S_{gf}=\frac{\kappa}{2}\int d^2\xi \left[\bar b_{-\alpha}\partial_+\theta^\alpha+\partial_-\bar\theta^\alpha b_{+\alpha}+\bar b_{-\alpha}\alpha^{\alpha\beta}b_{+\beta}\right]\nonumber\, ,\\
&&S_{FD}=\frac{\kappa}{2}\int d^2\xi \left[\bar C_\alpha\partial_+c^\alpha+(\partial_-\bar c^\alpha) C_{\alpha}\right]\, .
\end{eqnarray}
The Fadeev-Popov action is decoupled from the rest and, consequently, it can be omitted in further analysis.
On the equations of motion for $b$-fields
\begin{equation}
b_{+\alpha}=-(\alpha^{-1})_{\alpha\beta}\partial_+\theta^\alpha\, ,\quad \bar b_{-\alpha}=-\partial_-\bar\theta^\beta (\alpha^{-1})_{\beta\alpha}\, ,
\end{equation}
we obtain the final form of the BRST gauge fixed action
\begin{equation}
S_{gf}=-\frac{\kappa}{2}\int d^2\xi \partial_-\bar\theta^\alpha (\alpha^{-1})_{\alpha\beta}\partial_+\theta^\beta\, .
\end{equation}

\subsection{Fermionic T-duality}

We will perform fermionic T-duality using fermionic version of Buscher procedure similarly to Refs.\cite{nashi} where we worked without chiral gauge fixing. After introducing $S_{gf}$ the action still has a global shift
symmetry in $\theta^\alpha$ and $\bar\theta^\alpha$ directions. We introduce gauge fields $v^\alpha_\pm$ and
$\bar v^\alpha_\pm$ and replace ordinary world-sheet derivatives with covariant ones
\begin{equation}
\partial_\pm\theta^\alpha \to D_\pm \theta^\alpha\equiv\partial_\pm\theta^\alpha+v_\pm^\alpha\, , \quad \partial_\pm\bar\theta^\alpha \to D_\pm
\bar\theta^\alpha\equiv\partial_\pm\bar\theta^\alpha+\bar v_\pm^\alpha\, .
\end{equation}
In order to make the fields $v_\pm^\alpha$ and $\bar v_\pm^\alpha$ to be unphysical we add the following terms in the action
\begin{equation}
S_{gauge}(\vartheta,v_\pm,\bar \vartheta,\bar v_\pm)=\frac{1}{2}\kappa \int_\Sigma d^2\xi \bar \vartheta_\alpha (\partial_+
v_-^\alpha-\partial_- v^\alpha_+)+\frac{1}{2}\kappa \int_\Sigma d^2\xi  (\partial_+
\bar v_-^\alpha-\partial_- \bar v^\alpha_+)\vartheta_\alpha\, ,
\end{equation}
where $\vartheta_\alpha$ and $\bar\vartheta_\alpha$ are Lagrange multipliers.
The full gauge invariant action is of the form
\begin{equation}\label{eq:auxdejstvo}
S_{inv}(x, \theta,\bar\theta,\vartheta,\bar \vartheta,v_\pm,\bar v_\pm)=S(\partial_\pm x, D_- \theta, D_+ \bar\theta)+S_{gf}(D_- \theta, D_+ \bar\theta)+S_{gauge}(\vartheta,\bar \vartheta, v_\pm, \bar v_\pm)\, .
\end{equation}

Fixing $\theta^\alpha$ and $\bar\theta^\alpha$ to zero we obtain the gauge fixed action
\begin{eqnarray}\label{eq:gfa}
&&S_{fix}=\kappa \int_\Sigma d^2\xi \partial_+ x^\mu \left[\Pi_{+\mu\nu}+\frac{1}{2}\bar\Psi^\alpha_\mu(P^{-1})_{\alpha\beta}\Psi^\beta_\nu\right]\partial_-x^\nu+\frac{1}{4\pi}\int_\Sigma d^2\xi \Phi R^{(2)} \\ &{}& +\frac{\kappa}{2} \int_\Sigma \left[ \bar v_+^\alpha (P^{-1})_{\alpha\beta}v_-^\beta+\bar v_+^\alpha (P^{-1})_{\alpha\beta}\Psi^\beta_\nu\partial_-x^\nu+\partial_+x^\mu \bar\Psi^\alpha_\mu (P^{-1})_{\alpha\beta}v_-^\beta-\bar v_-^\alpha (\alpha^{-1})_{\alpha\beta}v_+^\beta\right]\nonumber \\ &{}& +\frac{\kappa}{2} \int_\Sigma d^2\xi \bar \vartheta_\alpha (\partial_+
v_-^\alpha-\partial_- v^\alpha_+)+\frac{\kappa}{2} \int_\Sigma d^2\xi  (\partial_+
\bar v_-^\alpha-\partial_- \bar v^\alpha_+)\vartheta_\alpha\, .\nonumber
\end{eqnarray}
Varying the above action with respect to
the Lagrange multipliers $\vartheta_\alpha$ and $\bar \vartheta_\alpha$ we obtain the initial action (\ref{eq:lcdejstvo}) because
\begin{equation}\label{eq:jed3}
\partial_+
v^\alpha_--\partial_- v^\alpha_+=0 \Longrightarrow v_\pm^\alpha=\partial_\pm \theta^\alpha \, ,\quad \partial_+
\bar v^\alpha_--\partial_- \bar v^\alpha_+=0 \Longrightarrow \bar v_\pm^\alpha=\partial_\pm \bar\theta^\alpha\, .
\end{equation}

The equations of motion for $v_\pm^\alpha$ and $\bar v_\pm^\alpha$ give
\begin{equation}\label{eq:jed1}
\bar v_-^\alpha=\partial_- \bar \vartheta_\beta \alpha^{\beta\alpha}\, ,\quad \bar v_+^\alpha=\partial_+ \bar \vartheta_\beta P^{\beta\alpha}-\partial_+ x^\mu \bar\Psi^\alpha_\mu\, ,
\end{equation}
\begin{equation}\label{eq:jed2}
v_+^\alpha=-\alpha^{\alpha\beta}\partial_+\vartheta_\beta\, ,\quad v_-^\alpha=-P^{\alpha\beta}\partial_-\vartheta_\beta-\Psi^\alpha_\mu \partial_- x^\mu\, .
\end{equation}
Substituting these expressions in the action $S_{fix}$
we obtain the fermionic T-dual action
\begin{eqnarray}\label{eq:tdualact}
&{}& {}^\star S(\partial_\pm x, \partial_- \vartheta, \partial_+
\bar \vartheta)=\kappa\int_\Sigma d^2\xi \partial_+ x^\mu  \Pi_{+\mu\nu}\partial_- x^\nu+\frac{1}{4\pi}\int_\Sigma d^2\xi\;\; {}^\star\Phi R^{(2)}\, ,\\ &{}&+\frac{\kappa}{2}\int_\Sigma d^2\xi\left[\partial_+\bar \vartheta_\alpha P^{\alpha\beta}\partial_-\vartheta_\beta -\partial_+x^\mu\bar\Psi^{\alpha}_{\mu} \partial_-\vartheta_\alpha+\partial_+\bar \vartheta_\alpha\Psi^{\alpha}_{\mu}\partial_- x^\mu-\partial_-\bar\vartheta_\alpha \alpha^{\alpha\beta} \partial_+\vartheta_\beta\right]\, .\nonumber
\end{eqnarray}
It should be in the form of the initial action (\ref{eq:lcdejstvo})
\begin{eqnarray}
&&{}^\star S=\kappa \int_\Sigma d^2 \xi \partial_+ x^\mu \left[{}^\star \Pi_{+\mu\nu}+\frac{1}{2}{}^\star \Psi^{\alpha\mu} ({}^\star P^{-1})^{\alpha\beta}{}^\star\Psi_{\beta\nu}\right]\partial_- x^\nu+\frac{1}{4\pi}\int_\Sigma d^2\xi {}^\star\Phi R^{(2)}\\
&+&\frac{\kappa}{2} \int_\Sigma d^2 \xi \left[\partial_+ \bar\vartheta_\alpha ({}^\star P^{-1})^{\alpha\beta}\partial_-\vartheta_\beta + \partial_+x^\mu ({}^\star\bar\Psi {}^\star P^{-1})_{\mu}^\alpha\partial_- \vartheta^\alpha+\partial_+\bar\vartheta_\alpha ({}^\star P^{-1}{}^\star\Psi)^{\alpha}_\mu\partial_-x^\nu\right]\nonumber\\ &-&\frac{\kappa}{2}\int d^2\xi \partial_- \bar\vartheta_\alpha ({}^\star \alpha^{-1})^{\alpha\beta}\partial_+\vartheta_\beta\, ,
\end{eqnarray}
and so we get
\begin{equation}\label{eq:Psidual}
{}^\star\Psi_{\alpha \mu}=(P^{-1}\Psi)_{\alpha\mu}\, ,\; {}^\star\bar\Psi_{\mu\alpha}=-(\bar\Psi P^{-1})_{\mu\alpha}\, ,
\end{equation}
\begin{equation}\label{eq:Fdual}
{}^\star P_{\alpha\beta}=(P^{-1})_{\alpha\beta}\, ,\quad {}^\star \alpha_{\alpha\beta}=(\alpha^{-1})_{\alpha\beta}\, .
\end{equation}
From the condition
\begin{equation}
{}^\star \Pi_{+\mu\nu}+\frac{1}{2}{}^\star \bar\Psi_{\alpha\mu}\,({}^\star P^{-1})^{\alpha\beta}\,{}^\star \Psi_{\beta\nu}=\Pi_{+\mu\nu}\, ,
\end{equation}
we read the fermionic T-dual metric and Kalb-Ramond field
\begin{eqnarray}\label{eq:GBdual}
&&{}^\star G_{\mu\nu}=G_{\mu\nu}+\frac{1}{2}\left[ (\bar\Psi P^{-1}\Psi)_{\mu\nu}+(\bar\Psi P^{-1}\Psi)_{\nu\mu}\right]\, ,\nonumber \\&&{}^\star B_{\mu\nu}=B_{\mu\nu}+\frac{1}{4}\left[ (\bar\Psi P^{-1}\Psi)_{\mu\nu}-(\bar\Psi P^{-1}\Psi)_{\nu\mu}\right]\, .
\end{eqnarray}
Dilaton transformation under fermionic T-duality will be presented in the section 4.
Let us note that two successive dualizations give the initial
background fields.

The T-dual transformation laws are connection between initial and T-dual coordinates. We can obtain them combining the different solutions of equations of motion for $v_\pm^\alpha$ and $\bar v_\pm^\alpha$ (\ref{eq:jed3}) and (\ref{eq:jed1})-(\ref{eq:jed2})
\begin{equation}\label{eq:tlaw}
\partial_-\theta^\alpha\cong-P^{\alpha\beta}\partial_-\vartheta_\beta-\Psi^\alpha_\mu \partial_- x^\mu\, ,\quad
\partial_+\bar\theta^\alpha\cong\partial_+ \bar \vartheta_\beta P^{\beta\alpha}-\partial_+ x^\mu \bar\Psi^\alpha_\mu\, ,
\end{equation}
\begin{equation}\label{eq:228}
\partial_+\theta^\alpha\cong -\alpha^{\alpha\beta}\partial_+\vartheta_\beta\, ,\quad \partial_-\bar\theta^\alpha\cong \partial_- \bar\vartheta_\beta \alpha^{\beta\alpha}\, .
\end{equation}
Here the symbol $\cong$ denotes the T-duality relation. From these relations we can obtain inverse transformation rules
\begin{equation}\label{eq:tlawi}
\partial_-\vartheta_\alpha\cong -(P^{-1})_{\alpha\beta}\partial_-\theta^\beta-(P^{-1})_{\alpha\beta}\Psi^\beta_\mu\partial_- x^\mu\, ,\quad \partial_+\bar\vartheta_\alpha\cong \partial_+\bar\theta^\beta (P^{-1})_{\beta\alpha}+\partial_+ x^\mu \bar\Psi^\beta_\mu (P^{-1})_{\beta\alpha}\, ,
\end{equation}
\begin{equation}\label{eq:230}
\partial_+ \vartheta_\alpha\cong -(\alpha^{-1})_{\alpha\beta}\partial_+\theta^\beta\, ,\quad \partial_- \bar\vartheta_\alpha\cong \partial_- \bar\theta^\beta (\alpha^{-1})_{\beta\alpha}\, .
\end{equation}

Note that without gauge fixing in subsection 2.2, instead expressions for $\bar v_-^\alpha$ and $v^\alpha_+$ (first relations of (\ref{eq:jed1}) and (\ref{eq:jed2})), we would have only constraints
on the T-dual variables, $\partial_- \bar\vartheta_\alpha=0$ and $\partial_+ \vartheta_\alpha=0$. Consequently, integration over $v_\pm^\alpha$ and $\bar v^\alpha_\pm$ would be singular and we would lose
the part of T-dual transformations (\ref{eq:228}) and (\ref{eq:230}).

\section{Fermionic T-dualization in fermionic double space}
\setcounter{equation}{0}

In this section we will extend the meaning of the double space. We will introduce double fermionic space adding to the fermionic coordinates, $\theta^\alpha$ and $\bar\theta^\alpha$,
the fermionic T-dual ones, $\vartheta_\alpha$ and $\bar\vartheta_\alpha$. Then we will show that fermionic T-dualization can be represented as permutation of the appropriate fermionic coordinates and their T-dual partners.

\subsection{Transformation laws in fermionic double space}

In the same way as the double bosonic coordinates were introduced \cite{Duff,sazdam,sazda}, we double both fermionic coordinate as
\begin{equation}
\Theta^A=\left(
\begin{array}{c}
\theta^\alpha\\
\vartheta_\alpha
\end{array}\right)\, ,\quad \bar\Theta^A=\left(
\begin{array}{c}
\bar\theta^\alpha\\
\bar\vartheta_\alpha
\end{array}\right)\, .
\end{equation}
Each double coordinate has 32 real components. In terms of the double fermionic coordinates the transformation laws, (\ref{eq:tlaw})-(\ref{eq:230}),
can be rewritten in the form
\begin{equation}\label{eq:tdlaw}
\partial_- \Theta^A\cong-\Omega^{AB}\left[{\cal F}_{BC}\partial_- \Theta^C+{\cal J}_{-B}\right]\, ,\quad \partial_+ \bar\Theta^A\cong\left[\partial_+\bar\Theta^C {\cal F}_{CB} +\bar{\cal J}_{+B}\right]\Omega^{BA}\, ,
\end{equation}
\begin{equation}\label{eq:tdlawadd}
\partial_+ \Theta^A\cong -\Omega^{AB}{\cal A}_{BC}\partial_+\Theta^C\, ,\quad \partial_- \bar\Theta^A\cong \partial_- \bar\Theta^C {\cal A}_{CB}\Omega^{BA}\, ,
\end{equation}
where "fermionic generalized metric" ${\cal F}_{AB}$ has the form
\begin{equation}
{\cal F}_{AB}=\left(
\begin{array}{cc}
(P^{-1})_{\alpha\beta} & 0\\
0 & P^{\gamma\delta}
\end{array}\right)\, ,
\end{equation}
and
\begin{equation}\label{eq:matrixA}
{\cal A}_{AB}=\left(
\begin{array}{cc}
(\alpha^{-1})_{\alpha\beta} & 0\\
0 & \alpha^{\gamma\delta}
\end{array}\right)\, .
\end{equation}
${\cal F}_{AB}$ is bosonic variable but we put the name fermionic because it appears in the case of fermionic T-duality.

The double currents, $\bar{\cal J}_{+ A}$ and ${\cal J}_{-A}$, are fermionic variables of the form
\begin{equation}
\bar{\cal J}_{+A}=\left(
\begin{array}{c}
(\bar\Psi P^{-1})_{\mu\alpha}\partial_+ x^\mu\\
-\bar\Psi^\alpha_\mu \partial_+ x^\mu
\end{array}\right)\, ,\quad {\cal J}_{-A}=\left(
\begin{array}{c}
(P^{-1}\Psi)_{\alpha\mu}\partial_- x^\mu\\
\Psi^\alpha_\mu \partial_- x^\mu
\end{array}\right)\, ,
\end{equation}
while the matrix $\Omega^{AB}$ is constant
\begin{equation}
\Omega^{AB}=\left(
\begin{array}{cc}
0 & 1\\
1 & 0
\end{array}\right)\, ,
\end{equation}
where identity matrix is $16\times 16$.
By straightforward calculation we can prove the relations
\begin{equation}\label{eq:sodd}
\Omega^2=1\, ,\quad \det {{\cal F}_{AB}}=1\, .
\end{equation}
Consistency of the transformation laws (\ref{eq:tdlaw}) produces
\begin{equation}
(\Omega{\cal F})^2=1\, ,\quad {\cal J}_-={\cal F}\Omega {\cal J}_-\, ,\quad \bar{\cal J}_+=-\bar{\cal J}_+ \Omega {\cal F}\, .
\end{equation}

\subsection{Double action}

It is well known that equations of motion of initial theory are Bianchi identities in T-dual picture and vice versa.
As a consequence of the identities
\begin{equation}
\partial_+ \partial_- \Theta^A-\partial_-\partial_+ \Theta^A=0\, ,\quad \partial_+ \partial_- \bar\Theta^A-\partial_-\partial_+ \bar\Theta^A=0\, ,
\end{equation}
known as Bianchi identities, and relations (\ref{eq:tdlaw}) and (\ref{eq:tdlawadd}), we obtain the consistency conditions
\begin{equation}\label{eq:tdual1}
\partial_+ ({\cal F}_{AB}\partial_- \Theta^B+J_{-A})-\partial_-({\cal A}_{AB}\partial_+\Theta^B)=0\, ,
\end{equation}
\begin{equation}\label{eq:tdual2}
\partial_- (\partial_+ \bar\Theta^B {\cal F}_{BA}+\bar J_{+A})-\partial_+(\partial_-\bar\Theta^B {\cal A}_{BA})=0\, .
\end{equation}
The equations (\ref{eq:tdual1}) and (\ref{eq:tdual2})  are equations of motion of the following action
\begin{eqnarray}{\label{eq:doubleS}}
&&S_{double}(\Theta,\bar\Theta)=\\
&=&\frac{\kappa}{2}\int d^2\xi \left[\partial_+\bar\Theta^A {\cal F}_{AB}\partial_-\Theta^B+\bar {\cal J}_{+A}\partial_-\Theta^A+\partial_+\bar\Theta^A {\cal J}_{-A}-\partial_-\bar\Theta^A {\cal A}_{AB}\partial_+\Theta^B+L(x)\right]\nonumber\, ,
\end{eqnarray}
where $L(x)$ is arbitrary functional of the bosonic coordinates.

\subsection{Fermionic T-dualization of type II superstring theory as permutation of fermionic coordinates in double space}

In order to exchange $\theta^\alpha$ with $\vartheta_\alpha$ and $\bar\theta$ with $\bar\vartheta_\alpha$, let us introduce the permutation matrix
\begin{equation}\label{eq:taumat}
{\cal T}^A{}_B=\left(
\begin{array}{cc}
0 & 1\\
1 & 0
\end{array}\right)\, ,
\end{equation}
so that double T-dual coordinates are
\begin{equation}\label{eq:simetry1}
{}^\star\Theta^A={\cal T}^A{}_B \Theta^B\, ,\quad {}^\star \bar\Theta^A={\cal T}^A{}_B \bar\Theta^B\, .
\end{equation}

We demand that T-dual transformation laws for double T-dual coordinates ${}^\star \Theta^A$ and ${}^\star\bar\Theta^A$ have the same form as for initial ones $\Theta^A$ and $\bar\Theta^A$ (\ref{eq:tdlaw}) and (\ref{eq:tdlawadd})
\begin{equation}
\partial_- {}^\star\Theta^A\cong-\Omega^{AB}\left[{}^\star{\cal F}_{BC}\partial_- {}^\star\Theta^C+{}^\star {\cal J}_{-B}\right]\, ,\quad \partial_+ {}^\star\bar\Theta^A\cong\left[\partial_+{}^\star\bar\Theta^C {}^\star{\cal F}_{CB} +{}^\star \bar{\cal J}_{+B}\right]\Omega^{BA}\, ,
\end{equation}
\begin{equation}
\partial_+ {}^\star\Theta^A\cong -\Omega^{AB}{{}^\star\cal A}_{BC}\partial_+{}^\star\Theta^C\, ,\quad \partial_- {}^\star\bar\Theta^A\cong \partial_- {}^\star\bar\Theta^C {{}^\star\cal A}_{CB}\Omega^{BA}\, .
\end{equation}
Then the fermionic T-dual "generalized metric" ${}^\star {\cal F}_{AB}$ and T-dual currents, ${}^\star \bar{\cal J}_{+ A}$ and ${}^\star {\cal J}_{-A}$, with the help
of (\ref{eq:simetry1}) and (\ref{eq:tdlaw}), can be expressed in terms of initial ones
\begin{equation}\label{eq:simetry2}
{}^\star {\cal F}_{AB}={\cal T}_A{}^C {\cal F}_{CD}{\cal T}^D{}_B\, ,\quad {}^\star\bar{\cal J}_{+ A}={\cal T}_A{}^B \bar{\cal J}_{+ B}\, ,\quad {}^\star {\cal J}_{- A}={\cal T}_A{}^B {\cal J}_{- B}\, .
\end{equation}
The matrix ${\cal A}_{AB}$ transforms as
\begin{equation}
{}^\star {\cal A}_{AB}={\cal T}_A{}^C {\cal A}_{CD}{\cal T}^D{}_B=({\cal A}^{-1})_{AB}\, .
\end{equation}
Note that, as well as bosonic case, double space action (\ref{eq:doubleS}) has global symmetry under transformations (\ref{eq:simetry1}) if the conditions (\ref{eq:simetry2}) are satisfied.

From the first relation in (\ref{eq:simetry2}) we obtain the form of the fermionic T-dual R-R background field
\begin{equation}\label{eq:exp1}
{}^\star P_{\alpha\beta}=(P^{-1})_{\alpha\beta}\, ,
\end{equation}
while from the second and third equation we obtain the form of the fermionic T-dual NS-R background fields
\begin{equation}
{}^\star \Psi_{\alpha\mu}=(P^{-1})_{\alpha\beta}\Psi^\beta_\mu\, ,\quad {}^\star \bar\Psi_{\alpha\mu}=-\bar\Psi^\beta_\mu (P^{-1})_{\beta\alpha}\, .
\end{equation}
The non singular  matrix $\alpha^{\alpha\beta}$ transforms as
\begin{equation}\label{eq:exp2}
({}^\star \alpha)_{\alpha\beta}=(\alpha^{-1})_{\alpha\beta}\, .
\end{equation}
The expressions (\ref{eq:exp1})-(\ref{eq:exp2}) are in full agreement with the relations (\ref{eq:Psidual}) and (\ref{eq:Fdual}) obtained by the standard fermionic Buscher procedure. Consequently,
we showed that permutation of fermionic coordinates defined in (\ref{eq:taumat}) and (\ref{eq:simetry1}) completely reproduces fermionic T-dual R-R and NS-R background fields.

\subsection{Fermionic T-dual metric ${}^\star G_{\mu\nu}$ and Kalb-Ramond field ${}^\star B_{\mu\nu}$}

The expression $\Pi_{+\mu\nu}+\frac{1}{2}\Psi^\alpha_\mu (P^{-1})_{\alpha\beta}\Psi^\beta_\nu$ appears in the action (\ref{eq:lcdejstvo}) coupled
with $\partial_\pm x^\mu$, along which we do not T-dualize. It is an analogue of $ij$-term of Refs.\cite{Hull,Hull2} where $x^i$ coordinates are not T-dualized, and $\alpha\beta$-term
in \cite{bosdouble} where fermionic directions are undualized.

Taking into account the form of the doubled action (\ref{eq:doubleS}) we suppose that term $L(x)$ has the form
\begin{equation}
L(x)=2\partial_+ x^\mu \left(\Pi_{+\mu\nu}+{}^\star\Pi_{+\mu\nu}\right)\partial_- x^\nu\equiv \mathcal L+{}^\star \mathcal L\, ,
\end{equation}
where $\Pi_{+\mu\nu}$ is defined in (\ref{eq:pipm}) and ${}^\star \Pi_{+\mu\nu}$ is fermionic T-dual which we are going to find.
This term should be invariant under T-dual transformation
\begin{equation}\label{eq:deltaF}
{}^\star  {\mathcal L}  = {\mathcal L}  + \Delta {\mathcal L} \, .
\end{equation}
Using the fact that two successive T-dualization are identity transformation, we obtain
\begin{equation}
{\mathcal L}  = {}^\star  {\mathcal L} + {}^\star \Delta  {\mathcal L}  \, .
\end{equation}
Combining last two relations we get
\begin{equation}\label{eq:uslovdl}
{}^\star \Delta  {\mathcal L}  = -  \Delta  {\mathcal L}  \, .
\end{equation}
If $\Delta  {\mathcal L}=2\partial_+ x^\mu \Delta_{\mu\nu} \partial_- x^\nu$, we obtain the condition for $\Delta_{\mu\nu}$
\begin{equation}\label{eq:uslovd}
{}^\star \Delta_{\mu\nu} = - \, \Delta_{\mu\nu} \, .
\end{equation}

Using the relations (\ref{eq:Psidual}) and (\ref{eq:Fdual}) we realize that, up to multiplication constant, combination
\begin{equation}
\Delta_{\mu\nu}=\bar\Psi^\alpha_\mu (P^{-1})_{\alpha\beta}\Psi^\beta_\nu\, ,
\end{equation}
satisfies the condition (\ref{eq:uslovd}).
So, we conclude that
\begin{equation}
{}^\star \Pi_{+\mu\nu}=\Pi_{+\mu\nu}+c \bar\Psi^\alpha_\mu (P^{-1})_{\alpha\beta}\Psi^\beta_\nu\, ,
\end{equation}
where $c$ is an arbitrary constant. For $c=\frac{1}{2}$ we obtain the relations (\ref{eq:GBdual}). So, in double space formulation the fermionic T-dual NS-NS background fields can be obtained up to an
arbitrary constant under assumption that two successive T-dualizations produce initial action.

\section{Dilaton field in double fermionic space}
\setcounter{equation}{0}

Dilaton field transformation under fermionic T-duality is considered \cite{ferdual}. Here we will discuss some new features of dilaton transformation under
fermionic T-duality as well as the dilaton field in fermionic double space.

Because the dilaton transformation has quantum origin we start with the path integral for the gauge fixed action given in Eq.(\ref{eq:gfa})
\begin{equation}
Z=\int d \bar v_+^\alpha d\bar v_-^\alpha d v_+^\alpha d v_-^\alpha d\bar\vartheta_\alpha d\vartheta_\alpha    e^{i\;  S_{fix} (v_\pm, \bar v_\pm, \partial_\pm\vartheta, \partial_\pm \bar\vartheta)}   \, .
\end{equation}
For constant background case, after integration over the fermionic gauge fields $\bar v_\pm^\alpha$ and $v_\pm^\alpha$, we obtain the generating functional $Z$ in the form
\begin{equation}
Z=\int d\bar\vartheta_\alpha d\vartheta_\alpha \det{\left[ (P^{-1}\alpha^{-1})_{\alpha\beta}\right]} e^{i\;{}^\star S(\vartheta,\bar\vartheta)}\, ,
\end{equation}
where ${}^\star S(\vartheta,\bar\vartheta)$ is T-dual action given in Eq.(\ref{eq:tdualact}). We are able to perform such integration thank to the facts that we fixed the gauge in subsection 2.2.

Note that here we multiplied with determinants of $P^{-1}$ and $\alpha^{-1}$ because we integrate over Grassman fields $v_\pm^\alpha$ and $\bar v_\pm^\alpha$. We can choose that $\det\alpha=1$, and the generating
functional gets the form
\begin{equation}
Z=\int d\bar\vartheta_\alpha d\vartheta_\alpha \det{\left[ (P^{-1})_{\alpha\beta}\right]} e^{i\;{}^\star S(\vartheta,\bar\vartheta)}\, .
\end{equation}
This produces the fermionic T-dual transformation of dilaton field
\begin{equation}\label{eq:fistar}
{}^\star \Phi=\Phi+\ln{\det{\left[(P^{-1})_{\alpha\beta}\right]}}=\Phi-\ln \det P^{\alpha\beta}\, .
\end{equation}

Let us calculate $\det P^{\alpha\beta}$ using the expression
\begin{equation}
(P{P_s^{-1}}P^T)^{\alpha\beta}=P_s^{\alpha\beta}-P_a^{\alpha\gamma}(P_s^{-1})_{\gamma\delta}P_a^{\delta\beta}\, ,
\end{equation}
where we introduce the symmetric and antisymmetric parts for initial background fields
\begin{equation}
P_s^{\alpha\beta}=\frac{1}{2}\left(P^{\alpha\beta}+P^{\beta\alpha}\right)\, ,\quad P_a^{\alpha\beta}=\frac{1}{2}\left(P^{\alpha\beta}-P^{\beta\alpha}\right)\, ,
\end{equation}
and similar expressions for T-dual background fields, ${}^\star P^s_{\alpha\beta}$ and ${}^\star P^a_{\alpha\beta}$.
Taking into account that
\begin{equation}
(P\cdot{}^\star P)^\alpha{}_\beta=\delta^\alpha{}_\beta\, ,
\end{equation}
we obtain
\begin{equation}
P_s^{\alpha\gamma}\;\;{}^\star P^s_{\gamma\beta}+P_a^{\alpha\gamma}\;\;{}^\star P^a_{\gamma\beta}=\delta^\alpha{}_\beta\, ,\quad P_s^{\alpha\gamma}{}^\star P^a_{\gamma\beta}+P_a^{\alpha\gamma}{}^\star P^s_{\gamma\beta}=0\, .
\end{equation}
From these two equations we obtain
\begin{equation}
{}^\star P^s_{\alpha\beta}=\left[(P_s-P_a P_s^{-1}P_a)^{-1}\right]_{\alpha\beta}\, ,
\end{equation}
and, consequently, we have
\begin{equation}
(PP_s^{-1}P^T)^{\alpha\beta}=\left[({}^\star P_s)^{-1}\right]^{\alpha\beta}\, .
\end{equation}
Taking determinant of the left and right-hand side of the above equation we get
\begin{equation}
\det P^{\alpha\beta}= \sqrt{\frac{\det P_s}{\det {}^\star P_s}}\, ,
\end{equation}
which produces
\begin{equation}
{}^\star \Phi=\Phi-\ln \sqrt{\frac{\det P_s}{\det {}^\star P_s}}\, .
\end{equation}
Using the fact that $P^{\alpha\beta}=e^{\frac{\Phi}{2}}F^{\alpha\beta}$ and ${}^\star P^{\alpha\beta}=e^{\frac{{}^\star \Phi}{2}}{}^\star F^{\alpha\beta}$, fermionic T-dual transformation law for dilaton takes the form
\begin{equation}
{}^\star \Phi=\Phi-\ln\sqrt{e^{8\left(\Phi-{}^\star\Phi\right)}\frac{\det F_s}{\det {}^\star F_s}}\, ,
\end{equation}
and finally we have
\begin{equation}
{}^\star \Phi=\Phi+\frac{1}{6}\ln\frac{\det F_s}{\det {}^\star F_s}\, .
\end{equation}
It is obvious that two successive T-dualizations act as identity transformation
\begin{equation}\label{eq:dualphi}
{}^\star{}^\star \Phi=\Phi\, .
\end{equation}

We can conclude that only symmetric parts of the R-R field strengths give contribution to the transformation of dilaton field under fermionic T-duality. In type IIA superstring theory R-R field strength $F^{\alpha\beta}$
contains tensors $F^A_0$, $F^A_{\mu\nu}$ and $F^A_{\mu\nu\rho\lambda}$, while in type IIB $F^{\alpha\beta}$ contains $F^B_\mu$, $F^B_{\mu\nu\rho}$ and self dual part of $F^B_{\mu\nu\rho\lambda\omega}$. Using
the conventions for gamma matrices from the appendix of the first reference in \cite{jopol} (see Appendix A), we conclude that symmetric part of $F^{\alpha\beta}$ in type IIA contains scalar $F^A_0$ and 2-rank tensor $F^A_{\mu\nu}$, while in
type IIB superstring theory it contains 1-rank $F^B_{\mu}$ and self dual part of 5-rank tensor $F^B_{\mu\nu\rho\lambda\omega}$.

Let us write the path integral for double action (\ref{eq:doubleS})
\begin{equation}
Z_{double}=\int d\Theta^A d\bar\Theta^A e^{iS_{double}(\Theta,\bar\Theta)}\, .
\end{equation}
Because $\det {\cal F}=1$ and $\det{\cal A}=1$ we obtain that dilaton field in double space is invariant under fermionic T-duality. Consequently, a new dilaton should be introduced (see \cite{sazdam,sazda}),
invariant under T-duality transformations. Because of the relation (\ref{eq:dualphi})
we define the T-duality invariant dilaton as
\begin{equation}
\Phi_{inv}=\frac{1}{2}\left({}^\star \Phi+\Phi\right)=\Phi+\frac{1}{12}\ln\frac{\det F_s}{\det {}^\star F_s}\, ,\quad {}^\star\Phi_{inv}=\Phi_{inv}\, .
\end{equation}

\section{Concluding remarks}
\setcounter{equation}{0}

In this article we considered the fermionic T-duality of the type II superstring theory using the double space approach. We used the action
of the type II superstring theory in pure spinor formulation neglecting ghost terms and keeping all terms up to the
quadratic ones which means that all background fields are constant.

Using equations of motion with respect to the fermionic momenta $\pi_\alpha$ and $\bar\pi_\alpha$ we eliminated
them from the action. We obtained the action expressed in terms of the derivatives $\partial_\pm x^\mu$, $\partial_-\theta^\alpha$ and
$\partial_+\bar\theta^\alpha$, where $\theta^\alpha$ and $\bar\theta^\alpha$ are fermionic coordinates. Because $\theta^\alpha$ appears in the action in the form $\partial_-\theta^\alpha$ and $\bar\theta^\alpha$ in the form $\partial_+\bar\theta^\alpha$,
there is a local chiral gauge symmetry with parameters depending on $\sigma^\pm=\tau\pm\sigma$. We fixed this gauge invariance using BRST approach.

Using the Buscher approach we performed fermionic T-duality procedure and obtained the form of the fermionic T-dual background fields. It is obvious that two successive fermionic T-dualization produces
initial theory i.e. they are equivalent to the identity transformation.

In the central point of the article we generalize the idea of double space and show that fermionic T-duality can be represented as permutation in fermionic double space.
In the bosonic case double space spanned by coordinates $Z^M=(x^\mu,y_\mu)$ can be obtained adding T-dual coordinates $y_\mu$ to the initial ones $x^\mu$. Using analogy with the bosonic case
we introduced double fermionic space doubling the initial coordinates $\theta^\alpha$ and $\bar\theta^\alpha$ with their fermionic T-duals, $\vartheta_\alpha$ and $\bar\vartheta_\alpha$. Double
fermionic space is spanned by the coordinates $\Theta^A=(\theta^\alpha,\vartheta_\alpha)$ and $\bar\Theta^A=(\bar\theta^\alpha,\bar\vartheta_\alpha)$.

T-dual transformation laws and their inverse ones are rewritten in fermionic double space by single relation
introducing the fermionic generalized metric ${\cal F}_{AB}$ and currents ${\cal J}_{- A}$ and $\bar{\cal J}_{+A}$.
Demanding that transformation laws for fermionic T-dual double coordinates, ${}^\star \Theta^A={\cal T}^A{}_B \Theta^B$ and ${}^\star \bar\Theta^A={\cal T}^A{}_B \bar\Theta^B$, are of the same form as those for $\Theta^A$ and $\bar\Theta^A$, we obtained fermionic T-dual
generalized metric ${}^\star {\cal F}_{AB}$ and currents ${}^\star{\cal J}_{- A}$ and ${}^\star\bar{\cal J}_{+A}$. These transformations act as symmetry transformations of the double action (\ref{eq:doubleS}).
They produce the form of the fermionic T-dual NS-R and R-R background fields which are in full accordance with the results obtained by standard
Buscher procedure.

The expressions for T-dual metric ${}^\star G_{\mu\nu}$ and Kalb-Ramond field ${}^\star B_{\mu\nu}$ cannot be found from double space formalism because they do not appear in the T-dual transformation laws.
These expressions, up to arbitrary constant, are obtained assuming that two successive T-dualization act as identity transformation.

We considered transformation
of dilaton field under fermionic T-duality. We derived the transformation law for dilaton field and concluded that just symmetric parts of R-R field strengths, $F_s^{\alpha\beta}$ and ${}^\star F^s_{\alpha\beta}$, affected
the dilaton transformation law. This means that in the case of type IIA scalar and 2-rank tensor have influence on the dilaton transformation, while in the case of type IIB 1-rank tensor and self-dual part of 5-rank tensor
take that role.

Therefore, we extended T-dualization in double space to the fermionic case. We proved that permutation of fermionic coordinates with corresponding T-dual ones in double space is equivalent to the fermionic T-duality
along initial coordinates $\theta^\alpha$ and $\bar\theta^\alpha$.
\appendix

\section{Gamma matrices}
\setcounter{equation}{0}

In the appendix of the first reference of \cite{jopol} one specific representation of gamma matrices is given. Here we will calculate the transpositions of
basis matrices $(C\Gamma_{(k)})^{\alpha\beta}$ for $k=1,2,3,4,5$, where $C$ is charge conjugation operator.

The charge conjugation operator is antisymmetric matrix, $C^T=-C$, and it acts on gamma matrices as
\begin{equation}
C\Gamma^\mu C^{-1}=-(\Gamma^\mu)^T\, .
\end{equation}
Now we have
\begin{equation}
(C\Gamma^\mu)^T=(\Gamma^\mu)^T C^T=-(\Gamma^\mu)^T C=C\Gamma^\mu C^{-1}C=C\Gamma^\mu\, ,
\end{equation}
\begin{equation}
(C\Gamma^\mu\Gamma^\nu)^T=C\Gamma^\mu\Gamma^\nu \Longrightarrow (C\Gamma^{[\mu\nu]})^T=C\Gamma^{[\mu\nu]}\, ,
\end{equation}
\begin{equation}
(C\Gamma^\mu\Gamma^\nu\Gamma^\rho)^T=-C\Gamma^\mu\Gamma^\nu\Gamma^\rho \Longrightarrow (C\Gamma^{[\mu\nu\rho]})^T=-C\Gamma^{[\mu\nu\rho]}\, ,
\end{equation}
\begin{equation}
(C\Gamma^\mu\Gamma^\nu\Gamma^\rho\Gamma^\lambda)^T=-C\Gamma^\mu\Gamma^\nu\Gamma^\rho\Gamma^\lambda \Longrightarrow (C\Gamma^{[\mu\nu\rho\lambda]})^T=-C\Gamma^{[\mu\nu\rho\lambda]}\, ,
\end{equation}
\begin{equation}
(C\Gamma^\mu\Gamma^\nu\Gamma^\rho\Gamma^\lambda\Gamma^\omega)^T=C\Gamma^\mu\Gamma^\nu\Gamma^\rho\Gamma^\lambda\Gamma^\omega \Longrightarrow (C\Gamma^{[\mu\nu\rho\lambda\omega]})^T=C\Gamma^{[\mu\nu\rho\lambda\omega]}\, .
\end{equation}

\end{document}